\documentclass[sigconf]{acmart}
\pdfoutput 1
\AtBeginDocument{%
  \providecommand\BibTeX{{%
    \normalfont B\kern-0.5em{\scshape i\kern-0.25em b}\kern-0.8em\TeX}}}

\setcopyright{acmcopyright}
\copyrightyear{2023}
\acmYear{2023}
\acmDOI{XXXXXXX.XXXXXXX}

\acmConference[ICSE'24]{46th International Conference on Software Engineering (ICSE 2024)}{April 14-20, 2024}{Lisbon, Portugal}
\acmBooktitle{ICSE'24:46th International Conference on Software Engineering (ICSE 2024) April 14-20, 2024, Lisbon, Portugal} 
\acmPrice{}
\acmISBN{}

\usepackage{enumitem}
\usepackage{hyperref}
\usepackage{multirow}
\usepackage{booktabs}
\begin{document}

\title{Teaching Digital Accessibility to Industry Professionals using the Community of Practice Framework: An Experience Report}

 
\author{P D Parthasarathy}\email{p20210042@goa.bits-pilani.ac.in}\orcid{0000-0002-8723-2407}
\affiliation{%
\institution{BITS Pilani, KK Birla Goa Campus}
\city{}\state{Goa}
\country{India}
\postcode{403726}
}
\author{Swaroop Joshi}\email{swaroopj@goa.bits-pilani.ac.in}\orcid{0000-0003-4536-2446}
\affiliation{%
 \institution{BITS Pilani, KK Birla Goa Campus}
 \city{}\state{Goa}
 \country{India}
 \postcode{403726}
}


\renewcommand{\shortauthors}{}


\newcommand{\hiddenauthor}[1]{\author{Anonymous Author#1}\email{anon#1@university.edu}
\affiliation{%
  \institution{Anon Institution}
  \city{Anon City}\state{Anon State}
  \country{Country}
  \postcode{12345}
}}

\newcommand{\pwd}{person with disabilities}
\newcommand{\pwds}{persons with disabilities}
\newcommand{\pc}{\%}
\newcommand{\cop}{Community of Practice}
\newcommand{\ac}{accessibility cohort}
\begin{abstract}
    Despite recent initiatives aimed at improving accessibility, the field of digital accessibility remains markedly behind contemporary advancements in the software industry as a large number of real world software and web applications continue to fall short of accessibility requirements.
A persisting skills deficit within the existing technology workforce has been an enduring impediment, hindering organizations from delivering truly accessible software products. 
This, in turn, elevates the risk of isolating and excluding a substantial portion of potential users. 
In this paper, we report lessons learned from a training program for teaching digital accessibility using the Communities of Practice (CoP) framework to industry professionals. 
We recruited 66 participants from a large multi-national software company and assigned them to two groups: one participating in a CoP and the other using self-paced learning.
We report experiences from designing the training program, conducting the actual training, and assessing the efficiency of the two approaches. 
Based on these findings, we provide recommendations for practitioners in Learnng and Development teams and educators in designing accessibility courses for industry professionals.

\end{abstract}

\ccsdesc[500]{Social and professional topics~Software engineering education}
\ccsdesc[300]{Human-centered computing~Accessibility}

\keywords{Accessibility (a11y), massive open online courses, communities of practice, Computing Education}

\newenvironment{myquote}%
  {\list{}{\leftmargin=0.1in\rightmargin=0.1in}\item[]}%
  {\endlist}

\maketitle
\section{Introduction}
The goal of accessibility is to create digital applications that do not exclude people with disabilities from using them. 
The WebAIM 2023 report~\citep{WebAIMWebAIMMillion2023} on accessibility states that 96.3\% of the world's top 1 million websites do not offer full accessibility. 
Global software giants and various governments emphasize developing accessible software \cite{appleincAppleHumanInterfacend, microsoftincMicrosoftDesignInclusive2016, googleincGoogleDevelopingAccessibilitynd,tatianaiskandarMakingFacebookCom2020, govtofindiaGIGWGuidelinesIndiannd, theusgovernmentITAccessibilityLaws2023}. 
However, according to the Teach Access report 2023~\citep{BridgingAccessibleTechnology2023}, 44\% of participants reported that their current staff didn't have the accessible skills to meet their organizations' goals. 
56\% reported that it was ``difficult or very difficult'' for their organization to find job candidates with accessibility skills. 
The limited accessibility skills among software professionals are largely attributed to the lack of coverage of accessibility topics in the Computer Science curriculum~\citep{bakerSystematicAnalysisAccessibility2020, patelWhySoftwareNot2020}.

While various approaches to teaching accessibility in academia are being reported (discussed in the next section), there is little reporting on training indistry professionals on these topics.
We developed an accessibility training program and administered it using the \cop\ (CoP) framework~\citep{wengerCommunitiesPracticeLearning1998}, which has been effective training of practitioners and professionals in a variety of fields.
But, to our knowledge, it has not been used for teaching accessibility so far.

Wenger's model comprises four interconnected elements: community, practice, meaning-making, and identity. In our program, all these elements are clearly present. 
We aim to foster a community of accessibility experts within software companies. The participants are actively involved in the practice of designing, developing, and testing accessible software, and are learning from experienced personnel in the process. 
As they engage in the practice, they are also developing an understanding of what accessibility software development processes consists of. Ultimately, these experiences contribute to a transformation in the identity of an `accessibility ally' that can support accessibility needs of their respective subteams within the organization.

We recruited 66 software industry professionals from a Large sized Multi-National Company (more than 50,000 employees) that offers both product and services for B2B and B2C (mainly offers products for other business but does offer some products for direct customers) for an in-house \ac.
The first author is employed with this organization as an accessibility expert.
36 out of these 66 participants were placed in a CoP group, based on the availability of their time they had indicated, while others learned at their own pace.
The cohort ran for 12 weeks in the winter of 2022-23.

The rest of this paper is organized as follows: Sec.~\ref{sec:relatedwork} briefly describes the existing literature. Sec.~\ref{sec:method} details the course design and learning objectives, and Sec.~\ref{sec:cohort} explains the implementation of the \ac. We present our findings in Sec.~\ref{sec:findings} and discuss the impliations and present our recommendations along with the limitations in Sec.~\ref{sec:discussion}. Finally, we conclude in Sec.~\ref{sec:conclusion}.
\section{Related Work}
\label{sec:relatedwork}

In the Fall of 2022, Teach Access surveyed its members, partners, and networks to gauge the state of accessibility within their respective organizations. According to the survey results, 81\% of respondents indicated that they augment their teams with external consultants and contractors and/or invest in supplementary external training~\citep{BridgingAccessibleTechnology2023}. This trend underscores the deficiency of in-house staff when it comes to possessing adequate accessibility skills to meet the needs of their organizations.

Academia has made several efforts to integrate accessibility into computer science (CS) curricula to train future software developers. \citeauthor{wallerIncludingAccessibilityUndergraduate2009}~\cite{wallerIncludingAccessibilityUndergraduate2009} propose a distinctive method of imparting accessibility knowledge to undergraduates, wherein accessibility is treated as an integral facet of design and development, seamlessly woven throughout the entire curriculum rather than being confined to a separate course. 

In computing education, instructors have adopted diverse approaches to teaching accessibility. These strategies span a broad spectrum, ranging from dedicated courses explicitly centered on accessibility and assistive technology~\cite{waldDesign10Credit2008,matauschAssistecUniversityCourse2006,kurniawanGeneralEducationCourse2010} to the infusion of accessibility topics into existing computing curricula.
The latter includes areas like software engineering~\cite{el-glalyTeachingAccessibilitySoftware2020,ludiIntroducingAccessibilityRequirements2007,rossAccessibilityFirstclassConcern2017},
introductory programming courses (CS1/CS2)~\cite{cohenAccessibilityIntroductoryComputer2005,jiaInfusingAccessibilityProgramming2021},
web development~\cite{freireAccessibilityWebMultimedia2013, rosmaitaAccessibilityNowTeaching2006,wangHolisticPragmaticApproach2012},
android app development~\cite{bhatiaIntegratingAccessibilityMobile2023} and  
artificial intelligence~\cite{tsengExplorationIntegratingAccessibility2022}.

These courses are thoughtfully designed to achieve a variety of key accessibility learning outcomes through conventional lecture-based and project-based teaching methodologies. These outcomes typically encompass the following objectives:
\begin{itemize}
    \item Fostering awareness about accessibility issues.
    \item Providing technical knowledge, including guidelines like the Web Content Accessibility Guidelines (WCAG 2.1).
    \item Developing a sense of understanding and compassion for people with disabilities.
    \item Highlighting the diverse career opportunities available within the accessibility domain.
\end{itemize}

\citeauthor{patelWhySoftwareNot2020}~\citep{patelWhySoftwareNot2020} conducted a survey involving 77 industry professionals and discovered that formal education had left them ill-equipped to tackle accessibility challenges throughout the software development lifecycle.
Meanwhile, \citeauthor{masuwa-morganIntroducingAccessOntoOntology2008}~\citep{masuwa-morganIntroducingAccessOntoOntology2008} suggests employing ontology to encapsulate accessibility requirements within user requirement documents to ensure their adherence and fulfillment.
Despite the growing body of literature on the accessibility skills gap among professionals~\citep{patelWhySoftwareNot2020,martinLandscapeAccessibilitySkill2022} and the perceptions of software engineers regarding accessibility~\citep{biAccessibilitySoftwarePractice2022, durduPerceptionWebsiteAccessibility2020}, it appears that there is a dearth of literature addressing the teaching of accessibility skills to software industry professionals.

\subsection{CoP applied in various fields}

We discussed the highlights of the \cop\ framework in the previous section. Here we review some applications of the framework in a variety of fields.
It has been utilized by researchers before to investigate how newcomers acquire the skills of a new profession. For example, \citeauthor{cuddapahUsingWengerCommunities2011}~\citep{cuddapahUsingWengerCommunities2011} used it to demonstrate how new teachers, through interaction with their peers, learn about teaching, establish their professional identity, and become part of the teaching community. 
Similarly, \citeauthor{jimenez-silvaCommunityPracticeTeacher2012}~\citep{jimenez-silvaCommunityPracticeTeacher2012} discussed how a community of teacher-learners influences the beliefs and perceptions of English-language teachers regarding their profession. 
\citeauthor{gilbuenaFeedbackProfessionalSkills2015}~\citep{gilbuenaFeedbackProfessionalSkills2015} showed how interactions with engineering design coaches help students gain a deeper understanding of engineering skills and insights into the functioning of disciplinary and industrial communities of practice, thereby enhancing their professional engineering competence. 
More recently, \citeauthor{pittersonImportanceCommunityFostering2020}~\citep{pittersonImportanceCommunityFostering2020} applied the CoP framework to explore the development of new engineering education researchers in a training program.
It has also been employed in software engineering training: \citeauthor{paasivaaraTeachingScrumMaster2021}~\citep{paasivaaraTeachingScrumMaster2021} used it to train software professionals for the scrum master role.
\section{Course Design}
\label{sec:method}

We now describe the course leaning objectives (LOs) and an outline of the modules used.

\subsection{Learning Objectives}\label{sec:los}

At a high level, the course aims to introduce the participants to accessibility fundamentals, design concepts, technical knowledge for developing accessible software and testing it for accessibility, and best practices for reporting accessibility issues.

\subsubsection{Disability and accessibility fundamentals}
After completing the course, the participants will be able to:
\begin{itemize}
  \item Identify web accessibility's scope and relevance,
  \item Illustrate the need for developing accessible software,
  \item Recognize digital accessibility features and barriers,
  \item Identify Challenges \pwds{} could face while interacting with a software product,
  \item Locate accessibility standards and guidelines.
\end{itemize}

\subsubsection{Designing for Accessibility}

By the end of the course, learners will be able to
\begin{itemize}
  \item Explain how design principles and practices affect accessibility,
  \item Exemplify how accessible design helps \pwds{} interact with software products,
  \item Compare Universal Design and Individual Accommodations model,
  \item Summarize different accessible design strategies for visual, information, multimedia, animation, and form design.
\end{itemize}

\subsubsection{Developing for Accessibility}

By the end of the course, learners will be able to
\begin{itemize}
  \item Identify key terms used in accessibility engineering such as shadow-DOM (Document Object Model), Accessible Rich Internet Applications (ARIA), etc.,
  \item Apply accessible markup,
  \item Replicate accessibility issues,
  \item Evaluate various automated and manual accessibility testing strategies.
\end{itemize}

\subsubsection{Documenting and Reporting}

By the end of the course, learners will be able to
\begin{itemize}
  \item Analyze the severity of accessibility bugs,
  \item Apply the appropriate reporting tool,
  \item Formulate a plan for addressing accessibility issues given a report.
\end{itemize}

\subsection{Course Outline} \label{sec:course-outline}

To achieve these learning objectives, we created a course spanning four modules.
The course borrows heavily from the W3C Web Accessibility Initiative's curricula on web accessibility\footnote{\url{https://www.w3.org/WAI/curricula/}} and effective topics from earlier studies in teaching accessibility to students \cite{palanTeachingInclusiveThinking2017}.

We describe each module below.

\subsubsection{Accessibility Basics Module}
This module offers a broader perspective on accessibility, emphasizing its place within the larger context of inclusion and diversity, with the goal of shifting the paradigm from viewing accessibility merely as a \emph{compliance} item to understanding the importance of enhancing the user experience for \pwds.
It encourages consideration of the \emph{why} behind accessibility, not just the \emph{how}.
It covered these topics:
\begin{itemize}
  \item Disabilities, Challenges, and Assistive Technologies:
        \begin{itemize}
          \item Theoretical Disability Models, Evaluating the Advantages and Limitations of their Fundamental Assumptions.
          \item Categories and Characteristics of Disabilities and Associated Barriers.
          \item Individual-Level Assistive Technologies and Adaptive Approaches for Permanent, Temporary, and Episodic Disabilities.
        \end{itemize}
  \item Disability Demographics and Statistical Data Patterns, Along with Their Significance.
  \item International Declarations and Conventions on Disability Rights.
  \item Accessibility Standards and Regulations to ICT (Information and Communications Technology).
  \item Business case for accessibility, including brand enhancement, market reach, innovation, and legal risk reduction.
  \item Essentials of Web Accessibility: Covering accessibility POUR principles (Perceivable, Operable, Understandable, and Robust) and basics of WCAG standards.
  \item Strategies to integrate ICT Accessibility across the company.
\end{itemize}

\subsubsection{Accessibility Design Module}
This module introduces universal design, emphasizing that the responsibility for accessibility extends beyond software engineers to include designers (and other stakeholders).
It underscores the importance of collaboration between software engineers and user experience (UX) designers to guarantee better accessibility.
The module delved into the following topics:

\begin{itemize}
  \item Individualized Accommodations vs Universal Design
  \item Principles of Universal Design
  \item WCAG - A designers' view:
        \begin{itemize}
          \item Visual Design with emphasis on usage of color, cues, styles and flexible layouts
          \item Information Design with emphasis on tables, naming and grouping, structure of text.
          \item Navigation Design delving with menus, site and page navigation, including shortcuts.
          \item Interaction Design with detailing of keyboard and gesture-based interactions.
          \item Images, Graphics and Form design with specific details to alt text, animations, error handling and notifications.
          \item Accessibility annotations to convey components and additional information to engineers.
        \end{itemize}
\end{itemize}

\subsubsection{Accessibility Development Module}
This module focuses on the engineering aspects of creating accessible software, such as using semantic HTML, shadow-DOM, and ARIA and WCAG 2.1. The topics covered are:
\begin{itemize}
  \item Developing Accessible Software
        \begin{itemize}
          \item Guidelines, principles, and techniques for meeting success criteria of WCAG.
          \item Semantic HTML, Standard controls, and building custom widgets using WAI-ARIA best practices.
          \item Coding for accessible Structures, Tables, Rich applications and Forms.
        \end{itemize}
  \item Accessibility issues
        \begin{itemize}
          \item Interoperability and compatibility issues.
          \item Debugging and replicating Accessibility issues.
          \item Accessibility quality assurance.
          \item Testing with assistive technologies.
          \item Testing tools for the web.
        \end{itemize}
\end{itemize}

\subsubsection{Accessibility Reporting Module}
This module addresses how to report accessibility issues and remediate them.
The topic covered in this module are:
\begin{itemize}
  \item Identification of severity of accessibility bugs and prioritization of bugs (Potential to cause harm, Task blocking, difficult to complete, low impact).
  \item Template and usage of ACR (Accessibility Conformance Report) and VPAT(Voluntary Product Accessibility Template)
  \item Providing suggestions and/or approaches for resolving software bugs based on criteria such as optimal solutions, broad usability, feasibility, the choice between fixing or redesigning, and the steps for implementing the solution.
\end{itemize}

\section{The Accessibility Cohort}
\label{sec:cohort}

We reached out to the employees within the organization via emails that explained the purpose and included basic information about the program.
It also contained a link to a Google Form to express interest to participate in the `accessibility cohort' along with some queries about the participants.
Participants were promised an internal certification after successful completion of the assessments in the cohort.
Two reminders were sent on the fifth and tenth working day after the initial emails.
Due to some organizational reasons, we focused on the employees in the Asia-Pacific region, which has 1,472 employees excluding the sales and marketing teams. Initial emails and reminders were sent to all of them.
Seventy employees expressed interest, but four withdrew due to various reasons.
Thus, the final number of participants in the accessibility cohort was 66.

The accessibility cohort was structured to span 12 weeks from December 2022 to February 2023, with an expected weekly commitment of about 2.5 hours.
Participants who indicated that they can commit more than 2.5 hours per week were assigned to the \cop\ group (CoP), while the others were assigned to the Self Paced (SP) group.

\subsection{Participant Demographics}

\begin{table}[tbhp]
  \caption{\centering Participant Profile Summary (N=66)}\label{tab:participantSummary}
  \centering
  \begin{tabular}{lll}
    \toprule
    \multirow{1}{*}{Cateogry}                & \multicolumn{1}{l}{SubCategory}        & \multicolumn{1}{l}{Count}
    \\\midrule
    \multirow{4}{*}{Job Title}               & \multicolumn{1}{l}{Software Engineer}  & \multicolumn{1}{l}{45}    \\
                                             & \multicolumn{1}{l}{UX Designer}        & \multicolumn{1}{l}{13}    \\
                                             & \multicolumn{1}{l}{Management}         & \multicolumn{1}{l}{6}     \\
                                             & \multicolumn{1}{l}{Enginnering Intern} & \multicolumn{1}{l}{2}     \\\midrule
    \multirow{4}{*}{Experience}              & \multicolumn{1}{l}{0-3 Years}          & \multicolumn{1}{l}{27}    \\
                                             & \multicolumn{1}{l}{4-7 Years}          & \multicolumn{1}{l}{17}    \\
                                             & \multicolumn{1}{l}{8-15 Years}         & \multicolumn{1}{l}{8}     \\
                                             & \multicolumn{1}{l}{15+ Years}          & \multicolumn{1}{l}{14}    \\\midrule
    \multirow{3}{*}{Accessibility Knowledge} & \multicolumn{1}{l}{Beginner}           & \multicolumn{1}{l}{33}    \\
                                             & \multicolumn{1}{l}{Intermediate}       & \multicolumn{1}{l}{28}    \\
                                             & \multicolumn{1}{l}{Expert}             & \multicolumn{1}{l}{5}     \\\midrule
    \multirow{2}{*}{Job Location}            & \multicolumn{1}{l}{India}              & \multicolumn{1}{l}{61}    \\
                                             & \multicolumn{1}{l}{Australia}          & \multicolumn{1}{l}{5}     \\\midrule
    \multirow{2}{*}{Gender}                  & \multicolumn{1}{l}{Men}                & \multicolumn{1}{l}{29}    \\
                                             & \multicolumn{1}{l}{Women}              & \multicolumn{1}{l}{37}    \\\midrule
    \multirow{3}{*}{Identify as a PWD}       & \multicolumn{1}{l}{Yes}                & \multicolumn{1}{l}{3}     \\
                                             & \multicolumn{1}{l}{No}                 & \multicolumn{1}{l}{52}    \\
                                             & \multicolumn{1}{l}{Prefer not to say}  & \multicolumn{1}{l}{11}    \\\bottomrule
  \end{tabular}
\end{table}

Most participants are software engineers with less than three years of industry experience. Half of the participants reported their accessibility knowledge is at the beginner level, while only five out of the 66 participants self-identified as accessibility experts. Only 3 identified as \pwds.
Table \ref{tab:participantSummary} summarizes the profiles of participants.

The gender distribution presents an interesting picture: while the general gender ratio in the organization is 3 men:2 women, about 25\% more women have participated in the cohort as compared to men.

\subsection{The two groups}

\subsubsection{Self Paced (SP) Group}

30 out of the 66 participants who indicated they can only spend about 2.5 hours per week as part of the \ac\ were assigned to this group.
They were expected to watch video lectures based on the modules as described in Section~\ref{sec:course-outline} made available via the organization's internal learning management system (LMS).
The LMS also featured additional resources such as links to open-source resources, including text and videos.
The LMS incorporated gamification elements, providing points and badges upon completion of various submodules.
Each submodule was to be completed in a week, and resources were carefully organized, ensuring it could be finished within a 2.5-hour timeframe.

The resources included:

\begin{itemize}
  \item Internally created videos and knowledge articles on each module by accessibility experts in the organization.
  \item A Coursera course titled \textit{An Introduction to Accessibility and Inclusive Design} by Univerisity of Illinois - Participants had to audit the course.
  \item The Udacity course titled \textit{Web Accessibility: Developing with Empathy} a free course offered in collaboration with Google.
  \item Resources sourced from Deque University (Free version).
  \item Blogs such as a11y-cats\footnote{\url{https://a11y-cats.com/}}, MaxAbility\footnote{\url{https://www.maxability.co.in/Blog/}}, and Digitala11y\footnote{\url{https://www.digitala11y.com/acccessibility-archives/}}.
\end{itemize}

To promote communication and address questions among participants, all members of group one were included in a Slack workspace, an online instant-messaging platform widely used for workplace communication.

\subsubsection{\cop\ (CoP) Group}\label{sec:cop}

The other 33 participants who said they could spend more than 2.5 hours per week were assigned to this group.
The CoP group focused on forming a community and engaging in practice.
A weekly 2-hour session was conducted every Wednesday.
Members at the workplace in India participated in-person at the on-campus telepresence room, while the members from Australia or those working from home joined online.
Each session was structured as follows:

\begin{itemize}
  \item Exploration (45 min): Participants who self-identified as having expert or intermediate knowledge of accessibility presented a topic from the modules as outlined in Section~\ref{sec:course-outline}.
  \item Experience Sharing (45 min): Members actively participated by sharing their experiences, insights, and knowledge related to the week's topic through presentations, discussions, case studies, or demonstrations.
  \item Practice (20 min): Members solved some exercises relevant to the week's topic individually or in groups. The exercises involved resolution of existing accessibility issues or the exploration of examples, both good and bad, available on the internet.
  \item Reflection (10 min): Members summarized the key takeaways from the session, discussed potential action items, and contemplated how the insights gained could be applied in their respective work or practices. Additionally, volunteers to present topics for the next sessions were selected, and a summary of the session was posted in a designated Slack channel.
\end{itemize}

\subsubsection{The GAAD 2023 Event}

An event were organized for the Global Accessibility Awareness Day (GAAD) by the organization on May 18, 2023, a few weeks after the \ac\ ended.
The event featured several components, including a visually impaired guest speaker who addressed the audience on how Generative AI is helping visually impaired individuals in their day-to-day tasks, and a \textit{bug-bash program} in which each participant resolved accessibility issues within their respective team's actual product, and interactive awareness activities such as crosswords and puzzles focused on accessibility.
All 66 \ac\ members attended this event.

\subsection{Assessment Instruments}

We used four instruments to assess the participants' understanding of accessibility and related concepts.

\subsubsection{Self-reported proficiency}

Before and after the \ac, participants in both groups self reported their accessibility proficiency on two questions:
\begin{itemize}
  \item[(a)] the level of their accessibility knowledge (\textsc{beginner}, \textsc{intermediate}, or \textsc{expert}) and
  \item[(b)] a 5-point \textsc{Agree}-\textsc{Disagree} Likert scale on the statement: ``I can act as an accessibility ally for my team.''
\end{itemize}

\subsubsection{Quizzes}

Participants in both groups were expected to answer three quizzes on the LMS, one each after the Accessibility Basics Module, Accessibility Design Module, and Accessibility Development Module.
These quizzes were administered as multiple-choice assessments, with a maximum score of 25 points within 30 minutes.
The quiz was auto-evaluated and the solution key was made available after the quiz.

\subsubsection{Accessibility Audit Assignment}

At the end of the cohort, we asked the participants to audit an internal, outdated, deprecated web application for accessibility issues.
The accessibility team of our organization had previously audited this web app and identified 30 accessibility issues.

To establish a baseline for comparison, at the start of the cohort, the participants were asked to `test' the app and report any issues they feel may impact how a PWD would interact with the app.

The end of the cohort assignment was to compile a detailed report highlighting the accessibility issues they identified, the corresponding WCAG guidelines that these issues pertained to, the priority level assigned to each issue, and the recommended solutions.
Additionally, participants were required to include evidence of the accessibility problems, such as screenshots or videos.
They were given two weeks to complete the assignment.
Organizers evaluated each report by applying the following 5-point rubric to each accessibility issue and provided detailed feedback via 1-on-1 in-person or online meets:
\begin{itemize}
  \item Recognition of each accessibility issue: 2 points.
  \item Corresponding WCAG guideline for each issue: 1 point.
  \item Accurate assignment of priority level (e.g., potential user harm, task hindrance, task complexity, low impact): 0.5 points.
  \item Precise documentation of the issue through screenshots and screen recordings: 0.5 points.
  \item Recommendation to fix the issue: 1 point
\end{itemize}
The application had 30 accessibility issues. Hence, the maximum score possible in this assignment was 150 points.



\section{Findings}
\label{sec:findings}

\subsection{Self-reported proficiency}

\begin{figure}[thbp]
  \centering
  \includegraphics[scale=0.31]{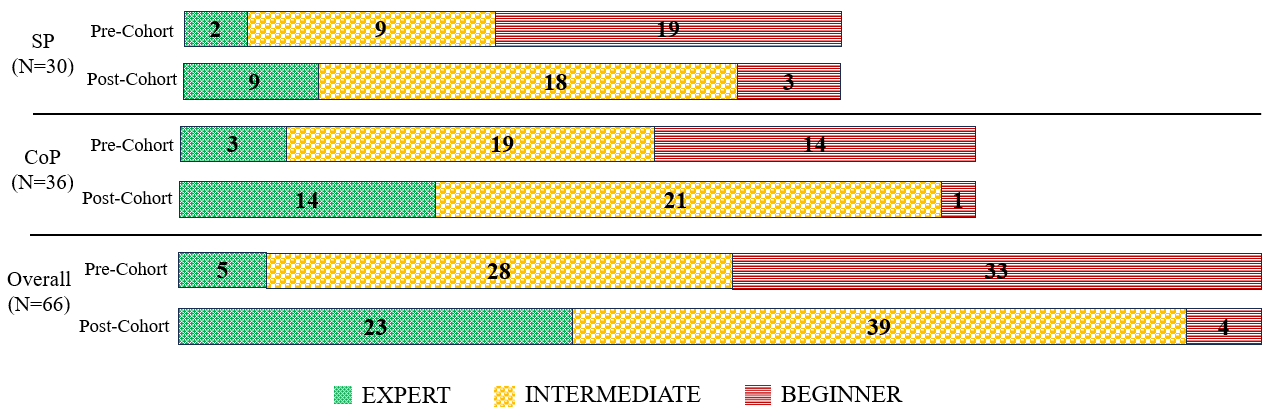}
  \caption{Accessibility expertise of the participants increased after the cohort}
  \Description[Participants' self-reported accessibility expertise]{Start of the cohort: 5 experts, 28 intermediate, 33 beginners. After the cohort: 23 experts, 39 intermediate, 4 beginners.}
  \label{fig:expertise}
\end{figure}

As expected, most participants report they have advanced to \textsc{intermediate} or \textsc{expert} level of accessibility knowledge after completing the \ac\ (Figure~\ref{fig:expertise}).
Only 4 self-report as \textsc{beginner}s at the end of the cohort, as against 33 at the start, while the number of \textsc{intermediate}s has gone up from 28 to 39 and the number of \textsc{expert}s from 5 to 23.

\begin{figure}[thbp]
  \centering
  \includegraphics[scale=0.31]{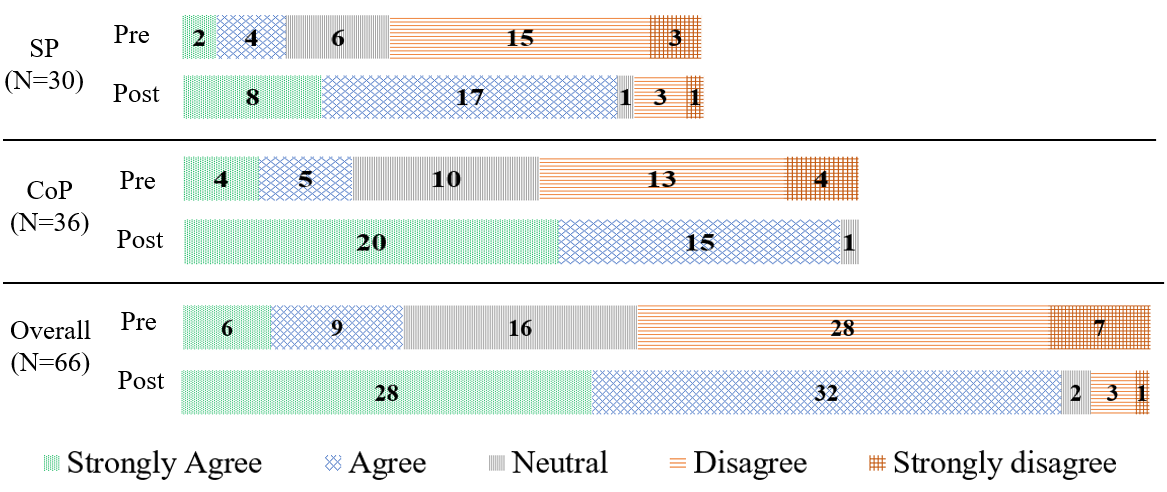}
  \caption{More than 90\% of the participants feel they can act as an accessibility ally for their teams}
  \Description[Participants' self reported confidence in being an accessibility ally for their team]{Start of the cohort: 6 strongly agreed, 9 agreed, 16 were neutral, 28 disagreed and 7 strongly disagreed. After the cohort: 28 strongly agred, 32 agreed, 2 were neutral, 3 disagreed and 1 strongly disagreed.}
  \label{fig:PrePostConfidence}
\end{figure}

Similarly, the confidence level of the participants to be able to help their team in accessibility matters has improved a lot (Figure~\ref{fig:PrePostConfidence}).
60 out of the 66 participants \textsc{agree} or \textsc{strongly agree} to the statement ``I can act as an accessibility ally for my team'' at the end of the cohort, as against 15 at the start.

\subsection{Quizzes}

\begin{figure}[tbhp]
  \centering
  \includegraphics[scale=0.39]{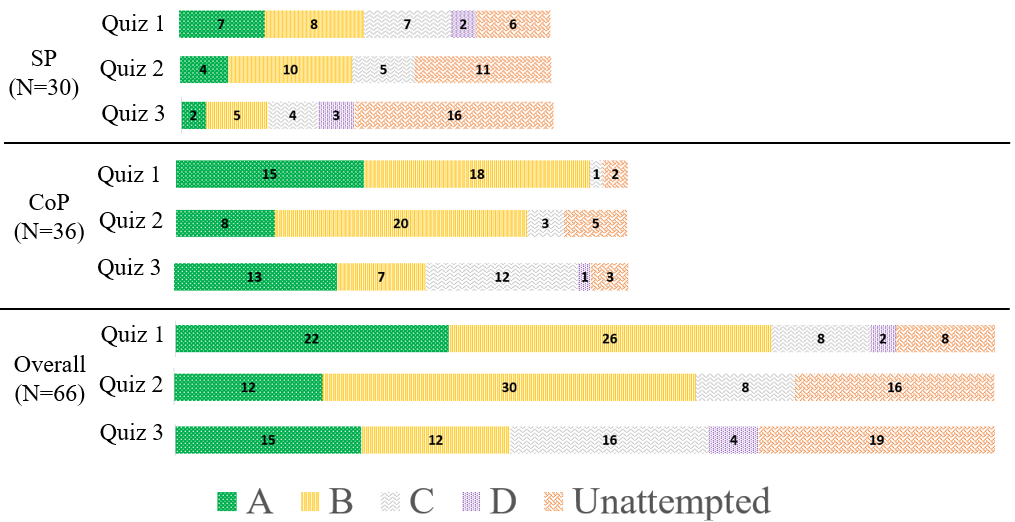}
  \caption{The CoP group has better performance on the quizzes compared to the SP group}
  \label{fig:QuizResults}
  \Description[Quiz Results]{13 participants from the CoP group got an A in Quiz 3, while only 2 from the SP group. Only 3 in CoP did not attempt Quiz 2, but 16 SP group members skipped it.}
\end{figure}

Each quiz was automatically evaluated on the LMS and one of the four letter grades was assigned.
Since the quizzes are of varying difficulty levels, scores should not be compared across quizzes.

However, across the two groups, SP and CoP, the latter shows better performance in all three quizzes.
Also, the number of participants who did not attempt the quizzes is comparitively low in the CoP group (2, 5, and 3) than in the SP group (6, 11, and 16).


\subsection{Accessibility Audit Assignment}

The assignment was evaluated for 150 points. For the baseline, the SP group (N=30) has a mean of 78.58 and standard deviation 19.91, while for the CoP group (N=36), mean is 75.12 and standard deviation 7.91.

First, we compared the baseline across the two groups.
Since the distribution is not normal (using the Shapiro-Wilk test), we applied the Mann-Whitney U test. We found that there is no significant difference in the baseline scores across the two groups.

In the SP group, the mean of the assignment score \emph{after} the cohort is 99.03 and standard deviation 20.78, while in the CoP group, 122 and 17.9, respectively.
We performed a gain-score analysis which is used to answer whether two groups differ in the gains on average~\citep{smolkowskiGainScoreAnalysis2019,joshiUsingAnonymityRoundsbased2017}.
The average gain (difference in the post and pre scores) in the SP group is 20.45 (std.\ dev.\ 13.3) and that in the CoP group is 46.88 (std.\ dev.\ 17.02).

Comparing the vectors of gains across two groups using Mann Whitney U test, we found there is a significant difference ($p<.05$).

\section{Discussion}
\label{sec:discussion}

\subsection{Effectiveness of the \cop\ framework}

Our findings suggest the CoP approach was more effective than the self-paced (SP) approach in various aspects of accessibility training.
A slightly higher number of participants in the CoP group (11 new, 14 total) believe their accessibility knowledge is at the \textsc{expert} level than in the SP group (7 new, 9 total).

One of the four important aspects of the CoP framework is \emph{identity}.
The second self-reporting question asked the participants to what degree do they agree believe they can act as an accessibility ally for their team, i.e., they can help the rest of their team develop or test software for accessibility.
All except one of the 36 CoP participants agreed or strongly agreed with the statement at the end of the cohort and no participant disagreed or strongly disagreed (whereas four participants in the SP group disagreed or strongly disagreed).
This suggests the CoP framework was effective in fostering the identity of the participants as accessibility professionals their teams could rely on.

More than half of the SP group did not attempt Quiz 3 (Figure~\ref{fig:QuizResults}), compared to only 3 out of the 36 in the CoP group did not attempt it.
Members of the SP group approached the cohort like a massive open online course (MOOC), which are effective in disseminating knowledge at scale, but are known to suffer with low completion rates~\citep{jordanInitialTrendsEnrolment2014,clowMOOCsFunnelParticipation2013}.
So it is likely that the members (partially) dropped out for similar reasons (they all completed the accessibility audit assignment though, which was required for getting the certification).
The CoP group, on the other hand, met regularly and engaged in activities as a \emph{community}, which may have led to better continued participation.
However, it should be noted that only those participants who had expressed they are available for more than 2.5 hours a week were assigned the CoP group, suggesting they were likely to be more motivated and available to participate in the \ac\ activities in the first place.

In the accessibility audit assignment, the participants demonstrated their application of the accessibility knowledge in practice by producing an audit report for a web application known to have accessibility issues.
A significantly better gain is observed in the CoP group compared to the SP group.
Through exploration, experience sharing, and reflection as described in Sec.~\ref{sec:cop}, the CoP members are likely to have improved both \emph{practice} and \emph{meaning making}, two other principles of COP.

\subsection{Participant Feedback}

We also collected open-ended feedback from the participants after completing the cohort.

Most Participants described that the topics covered were apt and lucid. One of the participants in the CoP group pointed out the extensive range of topics within the accessibility field with the following comment:
\begin{quote}
  [I] realised that what I knew before the course about accessibility was very little. Learned [a] lot of important pointers through the CoP. [I] [w]ill put them to use in future development.
\end{quote}

Another participant valued being part of the \emph{community}:
\begin{quote}
  I am grateful for the opportunity to be part of this community and would like to express my appreciation for the valuable learning and growth it has provided me. The resources and sessions were exceptional and catered to participants with diverse backgrounds and skill levels. I got to learn from others and they are a ping away to support me.
\end{quote}

A participant, who self-identified as a \pwd, expressed how she felt the CoP was inclusive:
\begin{quote}
  As a participant with cognitive differences, I've often felt left behind in training programs. This experience was different -- it was truly inclusive, and I felt like an equal participant and the informal setup made me feel at ease
\end{quote}

Some participants also suggested some improvements for a future offering of the cohort:
\begin{itemize}
  \item Need for Organization-specific examples: An SP participant wrote, ``Most of the resources featured general good/bad examples and were not specific to our [organization's] use cases.'' They recommended using specific examples from actual projects within the organization.
 \item Role-specific deep dive sessions: A CoP participant said, ``There can be extended, optional sessions exclusively for developers to deep dive [into] building new components like cards using accessibility in mind and how to develop AI-based automation for quick testing,'' and suggested such sessions can focus specifically on the best practices for the development phase.
  \item Frequent motivation (for the SP participants): An SP group member noted the GAAD event was useful in motivating them and suggested holding similar participative and hands-on exercises. They added, ``I learned more in those because I couldn't focus on [the video lectures] and did not participate after [the] 8th week.''
\end{itemize}

\subsection{Recommendations for practitioners}
\label{sec:recommendations}

Based on the discussion above, we present some recommendations for pracitioners interested in providing accessibility training to their employees:

\subsubsection{\cop\ as an effective framework for training} 

Beyond the improved gains in accessibility knowledge demonstrated by the quantitative analysis above, the CoP approach developed a sense of community and identity as `accessibility experts.'
Through anecdotal evidence from the Slack logs and conversations with colleagues and managers, we found that many CoP members continued their learning beyond the cohort through the network gained in the program.
CoP encourages customization, active engagement, social learning, practical application, continuous learning, and networking—ideal for industry professionals.
While self-paced learning has its advantages, CoP's informal and peer-focused approach can better suit (medium sized groups of) industry professionals.
We recommend considering CoP at least as an optional, additional approach, rather than offering self-paced learning alone.

\subsubsection{Role-specific accessibility training}

While accessibility is a shared concern, providing a one size fits all training approach has proven ineffective. 
In our experience, engineers were less motivated to delve into accessible design, while designers felt overwhelmed by the intricate technical aspects of accessibility development and testing.
After a general modules about fundamentals at the start, we recommend using separate modules tailored for separate roles.

\subsubsection{Organization Specific Examples over Generic ones} 

In our case, active participants in Communities of Practice (CoP) displayed a deeper understanding of accessibility. This can be attributed to their thorough discussions addressing accessibility challenges and solutions directly tied to their organization's unique context. Consequently, this led to improved performance in the final accessibility audit assessment and subsequent releases of their product. In contrast, participants in SP group grasped fundamental principles but encountered challenges in applying them practically due to the absence of context-specific discussion around accessibility bugs, including organizational design subsystems and engineering techniques. We recommend enhancing the effectiveness of accessibility training by customizing examples to align with real-world use cases from participants' own organizations. This approach yields more substantial and directly applicable results compared to the generic examples typically used in MOOCs to illustrate accessibility principles and guidelines.

\subsubsection{Emphasis on the \emph{Why} more than the \emph{How}}

Based on our observations, engineers often showed a keen interest in tools and techniques, focusing on the ``how'' aspect of achieving accessibility. However, this approach may fall short in intricate scenarios, as developers may not fully grasp the broader ``why'' behind the need for these techniques and who benefits from them. They tend to implement accessibility more effectively when they comprehend the rationale behind accessibility features and gain insight into how individuals with disabilities utilize them. Therefore, we recommend emphasizing the ``why'' aspects of accessibility alongside the ``how'', and encourage participants to engage with individuals with disabilities to better understand their lived experiences and perspectives on accessible software.

\subsubsection{Involve PwD in the training over Simulations} 

Previous research conducted in academic settings has shown that engaging with individuals with disabilities (PWDs) can significantly enhance participants' learning experiences, foster empathy, and boost motivation when it comes to accessibility topics, especially when compared to simulations~\cite{el-glalyPresentingEvaluatingImpact2020}.
Accessibility simulations in educational contexts have been found problematic by various researchers~\cite{tigwellNuancedPerspectivesDisability2021,bennettPromiseEmpathyDesign2019}.
In our specific case, the active involvement of three participants who identified as \pwds\ proved highly beneficial.
Their contributions during cohort design and their ongoing support within the Community of Practice were instrumental in dispelling misconceptions and providing valuable input. Therefore, we strongly recommend practitioners involve individuals with disabilities in their courses. By inviting PWDs to demonstrate how they utilize assistive technologies, the accessibility features that are crucial for them, and the barriers they encounter, participants can gain invaluable insights.

\subsubsection{Cover all disabilities} 

Aligned with the guidance provided by the W3C accessibility curriculum, we recommend practitioners take a comprehensive approach to accessibility training covering all disabilities. Our experience in curating materials highlighted an unintentional bias toward particular disability categories, such as an excessive emphasis on visual disabilities and screen reader usage, with limited coverage of cognitive disabilities. It's crucial to acknowledge that digital accessibility is important for individuals spanning a broad spectrum of disabilities, encompassing those related to hearing, cognition, learning, mobility, speech, and vision.

\subsubsection{Facililate participants in applying newly acquired accessibility skills} 

Motivate accomplished participants to pledge their commitment to accessibility advocacy, integrating it seamlessly into their yearly goals and personal growth plans. This dedication will equip them to consistently implement the skills they've gained in practical scenarios. Our firsthand experience demonstrated that CoP participants, who collectively embraced this commitment by pledging to convene regularly and sustain their involvement in accessibility matters, continued their engagement even beyond the cohort's conclusion. They actively participated in accessibility code reviews, addressed accessibility issues, and more. Conversely, MOOC participants often viewed their certification as a mere credential and lacked the motivation to put their acquired skills into practice.

\subsubsection{Usage of Games and Gamification for motivation}

To sustain participant interest and involvement, we employed crossword puzzles at regular intervals in our Slack channels and integrated them into our Global Accessibility Awareness Day (GAAD) event. Additionally, we introduced gamification badges throughout the cohort journey, rewarding participants for various achievements like attending sessions, completing MOOC modules, and contributing to accessibility issue resolutions. This approach garnered positive participant feedback, affirming its effectiveness. Therefore, we recommend utilizing games and gamification techniques to motivate participants whenever feasible.

Apart from the above, ensure that the course itself is accessible by making all materials and platforms used accessible, providing necessary support like captions and descriptions, and ensuring accessibility in physical venues, creating an inclusive learning experience for everyone.

\subsection{Limitations}

We report on our findings from an offering of a 12-week \ac\ for 66 employees in the Asia-Pacific region of a specific multi-national organization. The findings and recommendations should be interpreted with the limitations on generalizability in mind.
The cohort was organized into two groups, self-paced (30) and \cop\ (36), based on their self-reported availability.
Random assignments to groups could provide more sound findings. (Although, this being an experience report, the aim is not to establish whether CoP is a better approach to teach accessibility than self-paced learning.)

\section{Conclusion}
\label{sec:conclusion}

In this paper, we share insights gained from a training program that aimed to teach digital accessibility to industry professionals, employing Communities of Practice framework and comparing it against the self-paced learning that is commonly used. 
To our knowledge, this is the first report on teaching accessibility to industry professionals using \cop.
Our findings suggest the CoP framework was effective and appreciated by the participants.
We offer our lessons learned and recommendations to Learning and Development practitioners and educators for designing accessibility courses tailored to industry professionals.

\section*{Data Availability}
Data that support our findings are available as supplementary material and will be made available publicly upon acceptance on Zenodo.

\begin{acks}
  This work is supported by Birla Institute of Technology and Science, Pilani under grant GOA/ACG/2021-2022/Nov/05.
\end{acks}





\end{document}